\documentclass{webofc}
\usepackage[varg]{txfonts}   
\usepackage{gensymb,hyperref}
\newcommand{\eV}{\mathrm{eV}}
\newcommand{\EeV}{\mathrm{EeV}}
\newcommand{\Mpc}{\mathrm{Mpc}}
\newcommand{\km}{\mathrm{km}}
\newcommand{\yr}{\mathrm{yr}}
\newcommand{\sr}{\mathrm{sr}}
\newcommand{\PA}{\text{Auger}}
\newcommand{\TA}{\text{TA}}
\newcommand{\n}{\hat{\mathbf{n}}}
\newcommand{\TS}{\mathrm{TS}}
\newcommand{\dO}{\operatorname{d\Omega}}

\usepackage[dvipsnames]{xcolor}

\begin{document}
\title{2022 report from the Auger-TA working group on UHECR arrival directions}

\author{\firstname{A.}\ \lastname{di Matteo} \and
        \firstname{L.}\ \lastname{Anchordoqui} \and
        \firstname{T.}\ \lastname{Bister} \and
        \firstname{R.}\ \lastname{de~Almeida} \and
        \firstname{O.}\ \lastname{Deligny} \and
        \firstname{L.}\ \lastname{Deval} \and
        \firstname{G.}\ \lastname{Farrar} \and
        \firstname{U.}\ \lastname{Giaccari} \and
        \firstname{G.}\ \lastname{Golup} \and
        \firstname{R.}\ \lastname{Higuchi} \and
        \firstname{J.}\ \lastname{Kim} \and
        \firstname{M.}\ \lastname{Kuznetsov} \and
        \firstname{I.}\ \lastname{Mari\c{s}} \and
        \firstname{G.}\ \lastname{Rubtsov} \and
        \firstname{P.}\ \lastname{Tinyakov} \and
        \firstname{F.}\ \lastname{Urban}
        for the Pierre Auger 
        and Telescope Array 
        collaborations
}

\institute{Observatorio Pierre Auger, Av.\ San Mart{\'\i}n Norte 304, 5613 Malarg\"ue, Argentina \and
        Telescope Array Project, 201 James Fletcher Bldg, 115 S.\ 1400 East, Salt Lake City, UT 84112-0830, USA 
}

\abstract{%
After over 60 years, the powerful engines that accelerate ultra-high-energy cosmic rays (UHECRs) to the formidable energies at which we observe them from Earth remain mysterious. Assuming standard physics, we expect UHECR sources to lie within the local Universe (up to a few hundred Mpc). The distribution of matter in the local Universe is anisotropic, and we expect this anisotropy to be imprinted on the distribution of UHECR arrival directions. Even though intervening intergalactic and Galactic magnetic fields deflect charged UHECRs and can distort these anisotropies, some amount of information on the distribution of the sources is preserved. In this proceedings contribution, we present the results of the joint Pierre Auger Observatory and Telescope Array searches for (a) the largest-scale anisotropies (the harmonic dipole and quadrupole) and (b) correlations with a sample of nearby starburst galaxies and the 2MRS catalogue tracing stellar mass within~$250~\Mpc$. This analysis updates our previous results with the most recent available data, notably with the addition of 3~years of new Telescope Array data. The main finding is a correlation between the arrival directions of $12.1\%_{-3.1\%}^{+4.5\%}$~of UHECRs detected with $E \geq 38~\EeV$ by~Auger or with~$E \gtrsim 49~\EeV$ by~TA and the positions of nearby starburst galaxies on a ${15.1\degree}_{-3.0\degree}^{+4.6\degree}$~angular scale, with a $4.7\sigma$~post-trial significance, up from $4.2\sigma$ obtained in our previous study.
}
\maketitle

\section{Introduction}

Ultra-high-energy cosmic rays (UHECRs) are particles which impact the Earth's atmosphere with energies~$E \geq 1\,\EeV = 10^{18}\,\eV \approx 0.16\,\mathrm{J}$. Over 60 years after their discovery, we still do not know what are the sources of UHECRs and what is the mechanism that accelerates them to such formidable energies~\cite{AlvesBatista:2021gzc,Anchordoqui:2018qom}. To a smaller extent, we also still do not know what UHECRs are, namely what is their mass composition. What we do know is that the highest energy rays are most likely extra-Galactic. This can be inferred from the fact that their arrival direction distribution does not correlate with the matter density in the Milky Way~\cite{Abbasi:2016kgr,PierreAuger:2017pzq,PierreAuger:2021dqp}.

At the highest energies, UHECRs lose energy or disintegrate because of their interactions with cosmological background photons. This significantly reduces their propagation length, which is limited to a few hundred Mpc at a few tens of~$\EeV$ and rapidly decreases further at higher energies. At such distances the local Universe is highly anisotropic, the more so the closer we look. Therefore, we expect that the distribution of UHECR arrival directions is also anisotropic. However, charged UHECRs on their way to the Earth encounter intervening extragalactic and Galactic magnetic fields which deflect their trajectories, thereby obfuscating the true position of their sources, blurring and distorting their original anisotropy. The strength of magnetic deflections depends on \(ZB/E\), where \(Z\) is the UHECR electric charge, \(B\) the strength of the magnetic field, and \(E\) is the UHECR energy. Therefore, at higher energies we expect the anisotropic distribution of sources to be better preserved than at low energies because UHECRs travel more rectilinearly. Moreover, at higher energies the UHECR propagation horizon is smaller and the original anisotropy is more pronounced. However, at higher energies the number of UHECRs is also much smaller, thereby increasing the shot-noise contribution for anisotropy searches. Large-scale anisotropies, for example the harmonic dipole or quadrupole, corresponding to anisotropies on angular scales of $180\degree$ and $90\degree$, respectively, are expected to be much less affected by strong magnetic deflections of even some tens of degrees. On the other hand, associating events with specific sources is very hard unless the magnetic deflections, particularly the stochastic ones, are small enough. It is therefore fruitful to try to detect anisotropies at different energy scales.

The two largest cosmic-ray detector arrays in the world are the Pierre Auger Observatory (Auger) in Argentina and the Telescope Array (TA) in Utah, USA. Neither observatory can see the full sky: TA observes the full northern celestial hemisphere plus declinations~$\delta > -15.7\degree$ below the equator; Auger observes the southern hemisphere plus northern declinations up to~$\delta < +44.8\degree$. Without full-sky coverage, it is not possible to measure the dipole and quadrupole unless assumptions about higher multipoles are made. Moreover, there are comparably bright SBGs in both hemispheres, and with partial-sky coverage we would miss some of them. It is therefore crucial to combine the data from both observatories in order to maximize their data yield. This is the motivation behind the establishment of the TA--Auger working group on UHECR arrival directions.

The first, and thus far only, anisotropy detected with more than $5\sigma$ significance was a~$6.5\%$ dipole at~$E \geq 8\,\EeV$ reported by Auger~\cite{PierreAuger:2017pzq} with approximately $32\,000$~events, which has since grown in amplitude to~$7.3\%$ and in significance to~$6.6\sigma$~\cite{PierreAuger:2021dqp} (with about $44\,000$~events). Hints of a medium-scale anisotropy have been reported as a correlation between the arrival directions of UHECRs and the positions of nearby starburst galaxies (SBGs) on a ${15.5\degree}_{-3.2\degree}^{+5.3\degree}$~angular scale, disfavouring isotropy at $4.2\sigma$~post-trial significance, using a joint Auger-TA dataset~\cite{TelescopeArray:2021gxg} (for the Auger-only analysis see~\cite{PierreAuger:2018qvk,PierreAuger:2021rfz}).

\section{The data sets}
The Pierre Auger Observatory~\cite{PierreAuger:2015eyc} is a hybrid detector of UHECRs located in the Southern hemisphere in Argentina at a latitude of $-35.2^\circ$. It consists of a surface array of 1660 water-Cherenkov detectors covering an area of approximately 3000~km$^2$, overlooked by the fluorescence detector composed of 27 fluorescence telescopes. The detector has been taking data since January 2004. In this work, we use the dataset described in Ref.~\cite{PierreAuger:2022axr}, consisting of events detected by the surface detector array from 2004 Jan~01 to 2020 Dec~31. The geometrical exposure is $95\,700~\km^2~\yr~\sr$ for Auger vertical events (zenith angles~$\theta < 60\degree$) and $26\,300~\km^2~\yr~\sr$ for Auger inclined events ($60\degree \le \theta < 80\degree$). This corresponds to $39\,691$~events with $E_\PA\geq 8.53\,\EeV$ and $2635$~events with $E_\PA\geq 32\,\EeV$.

The Telescope Array~\cite{TelescopeArray:2012uws} is a hybrid detector of UHECRs located in the Northern hemisphere in Utah, USA at a latitude of $39.3^\circ$. It is taking data since May 2008. The surface detector of TA consists of 507 plastic scintillator detectors covering an area of about 700~km$^2$. The fluorescence detector of TA is composed of 38 fluorescence telescopes arranged in 3 towers overlooking the surface detector area. In this work, we use the events with the zenith angles $\theta<55^\circ$ detected by the TA surface detectors array from 2008 May~11 to 2022 May~10, namely three years more compared to the previous analyses~\cite{TelescopeArray:2021gxg,TelescopeArray:2021ygq}. The effective exposure is $18\,000~\km^2~\yr~\sr$ for TA events: this corresponds to $6014$~events with $E_\TA\geq 10\,\EeV$ and $395$~events with $E_\TA\geq 40.5\,\EeV$.

\section{The cross-calibration}

UHECR energy measurements are affected by sizeable systematic uncertainties ($\pm 14\%$ for Auger, $\pm 21\%$ for TA). If not corrected, a mismatch between energy scales can yield a spurious dipole. For instance, imagine that events with~$E_\text{true} = 10\,\EeV$ are reconstructed as~$E_\text{rec} = 9\,\EeV$ by Auger and as~$E_\text{rec} = 11\,\EeV$ by TA. If we collect all events with $E_\text{rec} \ge 10\,\EeV$, then events with $E_\text{true} = 10\,\EeV$ are included in the TA data set but not in the Auger one; therefore, the flux would appear to be larger in the North compared to the South. It is therefore necessary to cross-calibrate the two data sets before searching for anisotropies. This can be done from the data themselves without any extra assumptions on the nature of the energy determination systematics by comparing the data of the two observatories in the equatorial band where their exposures overlap~\cite{TelescopeArray:2014ahm}.

\begin{figure}[h]
\centering
\includegraphics[width=0.48\textwidth]{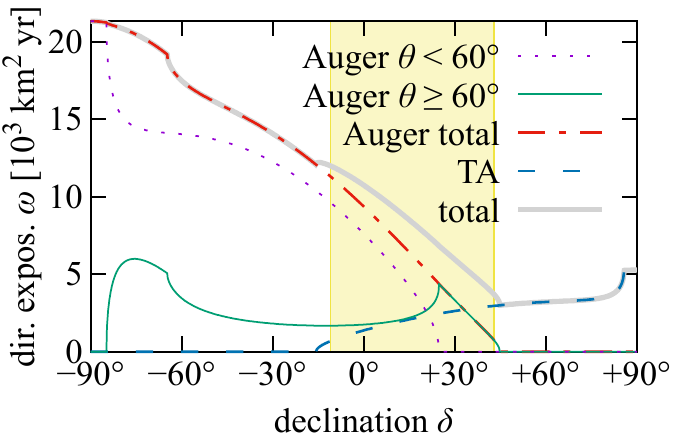}
\hspace{0.02\textwidth}%
\includegraphics[width=0.48\textwidth]{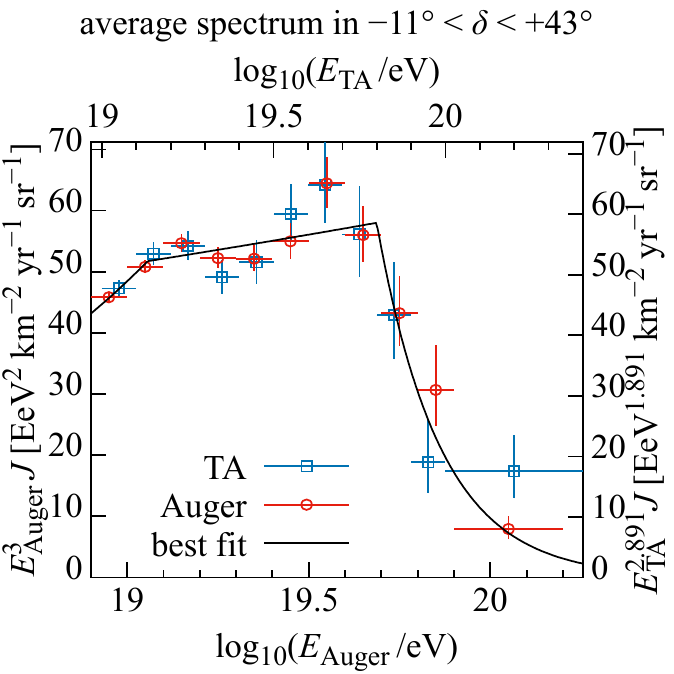}
\caption{Left:~the directional exposures of the datasets, with the declination band used for the energy cross-calibration highlighted.  Right:~the energy cross-calibration fit ($\chi^2/n_\text{dof} = 20.7/14$; $p = 0.11$).}
\label{fig:crosscal}
\end{figure}

In this contribution we follow the recipe outlined in~\cite{TelescopeArray:2021ygq} (which itself follows the earlier works~\cite{TelescopeArray:2014ahm,diMatteo:2018vmr,PierreAuger:2019oxg}) and update it using the latest available data sets. We simultaneously fit a twice-broken power-law model for the energy spectrum and a power-law model for the $E_\PA \leftrightarrow E_\TA$~conversion to the data in a common declination band as shown in figure~\ref{fig:crosscal}:
\begin{align*}
  E_\PA & = \hat E e^\alpha (E_\TA/\hat E)^\beta~~\text{, and}~~E_\TA = \hat E e^{-\alpha/\beta} (E_\PA/\hat E)^{1/\beta} \,,
\end{align*}
where $\hat E = 10\,\EeV$, $\alpha = -0.159 \pm 0.012$ and $\beta = 0.945 \pm 0.016$. The fit returns $\chi^2/n = 20.7/14$ with $n=14$ degrees of freedom, corresponding to a $p$-value of $0.11$. Note that in this fit we used data sets optimized for anisotropy studies at $E_\TA \ge 10\,\EeV$: the result must not be extrapolated to lower energies or used outside of the scope of this analysis.
    
\section{Latest results}

\subsection{Large-scale anisotropies: dipoles and quadrupoles}

The flux~$\Phi(\n)$ of UHECRs as a function of the arrival direction~$\n$ can be expanded into spherical harmonics $Y_{\ell m}(\n)$ as
\begin{align*}
\Phi(\n) &= \sum_{\ell=0}^{+\infty} \sum_{m=-\ell}^{+\ell} a_{\ell m} Y_{\ell m}(\n) 
          = \Phi_\text{avg} \times \left( 1 + \mathbf{d} \cdot \n + \frac{1}{2} \n \cdot \textsf{Q} \n + \cdots \right) \,,
\end{align*}
where the coefficients $a_{\ell m}$ represent anisotropies on scales~$\mathcal{O}(180\degree/\ell)$. The contribution of the two lowest non-trivial harmonics $\ell=1,2$, the dipole and quadrupole, can be rewritten in terms of a dipole vector $\mathbf{d}$ and the symmetric traceless quadrupole tensor $Q_{ij}$ as $\mathbf{d}=\sqrt{3} (a_{1{}1}\hat{\mathbf{x}} + a_{1{}-1}\hat{\mathbf{y}} + a_{1{}0} \hat{\mathbf{z}})/a_{0\/0}$,
and $Q_{xx}-Q_{yy} = 2\sqrt{15} a_{22}/a_{00}$, $Q_{xz} = \sqrt{15} a_{21}/a_{00}$, $Q_{yz} = \sqrt{15} a_{2-1}/a_{00}$, $Q_{zz} = 2\sqrt{5} a_{20}/a_{00}$, $Q_{xy} = \sqrt{15} a_{2-2}/a_{00}$ (the remaining components of $Q_{ij}$ are obtained from its symmetry and zero trace conditions).

The dipole amplitude~$\left|\mathbf{d}\right|$ and the quadrupole amplitude~$\left|\textsf{Q}\right|$
are relatively insensitive to magnetic fields, providing some information about sources: coherent deflections mainly affect the directions of $\mathbf{d},\textsf{Q}$ rather than their amplitudes, and small-scale random deflections only attenuate the $2^\ell$-pole by a factor~$\mathcal{O} \left( \exp\bigl(-\ell^2 \theta_\text{turb}^2\big/2\bigr)\right)$, where $\theta_\text{turb}$ is the average deflection accumulated from the scattering in turbulent magnetic fields, so that most of the~$\left|\mathbf{d}\right|$ and a sizeable fraction of the~$\left|\textsf{Q}\right|$ should survive (see also ref.~\cite{Eichmann:2020adn}).

Using a full-sky data set wherein the exposure $\omega(\n)$ is non-zero everywhere, each~$a_{\ell m}$ can be independently estimated in a way that is unbiased regardless of any assumption made about higher multipoles \cite{TelescopeArray:2021ygq}, as $\hat a_{\ell m} = \sum_k Y_{\ell m}(\n_k)/\omega(\n_k)$ where the sum runs over all $k$ UHECR events.

Our results are listed in table~\ref{tab-1}: we do not find any deviations from isotropy other than a weakly energy-dependent dipole towards a direction far away from the Galactic Center and a hint of a quadrupole at the highest energies roughly along the Supergalactic Plane. Compared to the previous analysis~\cite{TelescopeArray:2021ygq} the dipole component $d_y = 5.0\%\pm1.1\%\pm0.0\%$ has become more prominent, viz.\ $d_y = 4.8\%\pm1.1\%\pm0.0\%$, while most uncertainties have slightly decreased. The reconstructed strength of large-scale anisotropies remains at the low edge of a range of model expectations such as those of refs.~\cite{diMatteo:2017dtg,Ding:2021emg}, suggesting a medium to heavy mass composition.

\begin{table}
\centering
\caption{Our measurements of the dipole and quadrupole moments.  The first uncertainty is statistical, the second is due to the uncertainty in the cross-calibration of energy scales.  Values in \textit{italics (\textbf{bold})} are locally significant at $\ge 2\sigma$ ($\ge 4\sigma$).}
\label{tab-1}
	\begin{tabular}{@{}c|ccc@{}}
		\(E_\text{Auger}\) [EeV] & $[8.53, 16)$ & $[16, 32)$ & $[32, +\infty)$ \\
		\(E_\text{TA}\) [EeV] & $[10, 19.49)$ & $[19.49, 40.5)$ & $[40.5, +\infty)$ \\
		\hline
		$d_x~[\%]$ & $-0.2\pm1.1\pm0.0$ & $+0.9\pm1.9\pm0.0$ & $-4.4\pm\phantom{0}3.7\pm0.1$ \\
		$d_y~[\%]$ & \boldmath{$\it+5.0\pm1.1\pm0.0$} & $\it+4.4\pm1.9\pm0.0$ & $\it+10.0\pm\phantom{0}3.5\pm0.0\phantom{0}$ \\
		$d_z~[\%]$ & $-3.0\pm1.3\pm1.2$ & $-8.4\pm2.2\pm1.3$ & $+3.3\pm\phantom{0}4.4\pm3.5$ \\
		\hline
		$Q_{xx}-Q_{yy}~[\%]$ & $-4.3\pm4.6\pm0.0$ & $+12.9\pm8.1\pm0.0\phantom{0}$ & $\it+39.7\pm15.0\pm0.0\phantom{0}$ \\
		$Q_{xz}~[\%]$ & $-2.7\pm2.7\pm0.0$ & $+4.1\pm4.7\pm0.0$ & $+4.9\pm\phantom{0}9.7\pm0.1$ \\
		$Q_{yz}~[\%]$ & $-4.3\pm2.7\pm0.0$ & $-8.3\pm4.6\pm0.1$ & $+12.8\pm\phantom{0}9.1\pm0.3\phantom{0}$ \\
		$Q_{zz}~[\%]$ & $+0.5\pm3.1\pm1.5$ & $+4.5\pm5.4\pm1.5$ & $+22.0\pm10.3\pm4.1\phantom{0}$ \\
		$Q_{xy}~[\%]$ & $+1.3\pm2.3\pm0.0$ & $-0.6\pm4.0\pm0.1$ & $+4.0\pm\phantom{0}7.8\pm0.1$ \\
		\hline
	\end{tabular}
\end{table}

\begin{figure}[h]
\centering
\includegraphics[width=0.33\textwidth]{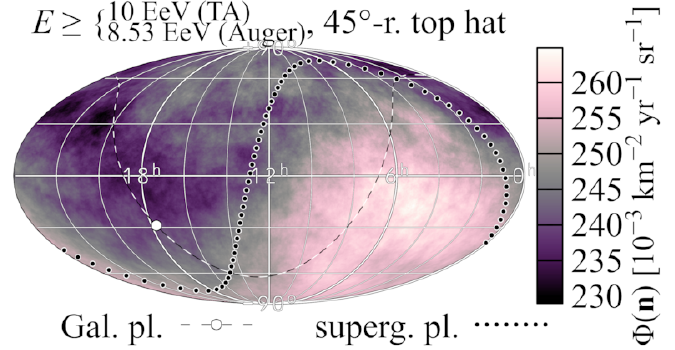}%
\includegraphics[width=0.33\textwidth]{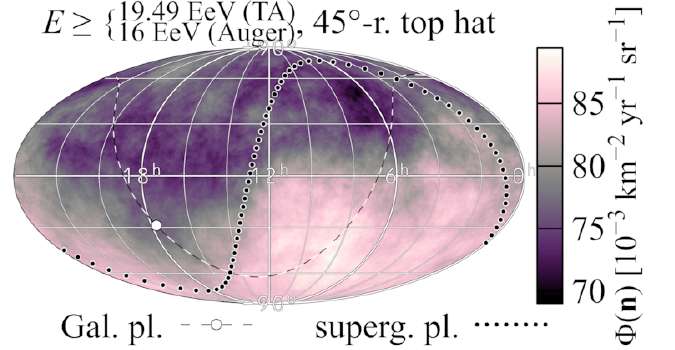}%
\includegraphics[width=0.33\textwidth]{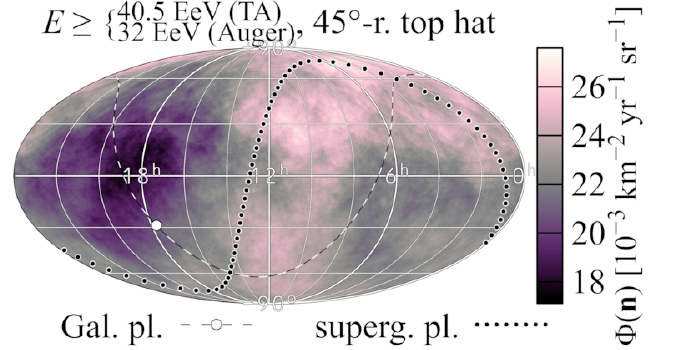}\\
\vspace{6pt}
\includegraphics[width=0.33\textwidth]{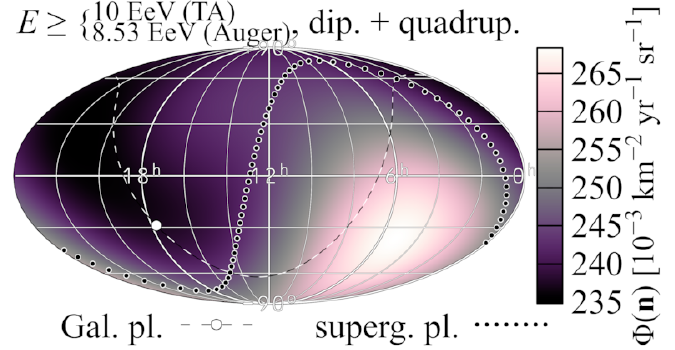}%
\includegraphics[width=0.33\textwidth]{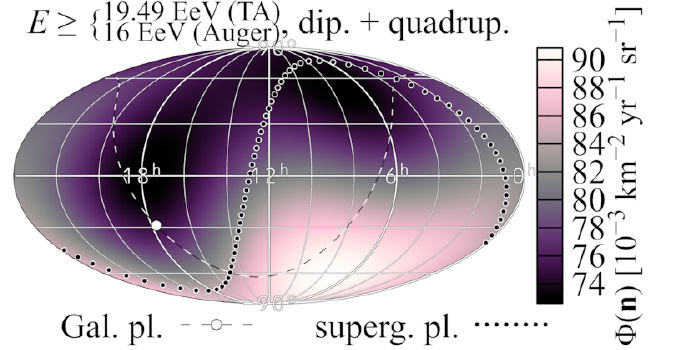}%
\includegraphics[width=0.33\textwidth]{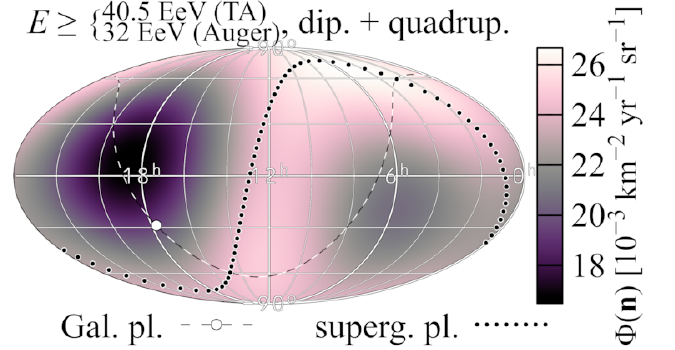}%
\caption{The observed flux in the three energy bins used for the large-scale anisotropies (top panels), and the reconstructed directional flux truncated to~$\ell \le 2$ (bottom panels).}
\label{fig:multi}
\end{figure}

\subsection{Medium-scale anisotropies: correlations with nearby galaxies}

In order to search for medium-scale anisotropies we need to focus on the highest energies, where the magnetic deflections are expected to be smaller and the UHECR propagation horizon shrinks (thus, fewer candidate sources remain, making the anisotropy from the distribution of the sources more pronounced). However, the amount of statistics available at those energies is dramatically lower, making ``blind'' searches very unlikely to succeed. We therefore perform targeted searches based on two different catalogues, following our previous study~\cite{TelescopeArray:2021gxg}. The first is a list of~$44\,113$ galaxies of all types at distances~$1~\Mpc \le D < 250~\Mpc$, based on the 2MASS catalogue with distances from HyperLEDA and weights assumed proportional to the near-infrared flux in the $K$-band ($2.2$~\textmu{m}). The second is a list of 44 starburst galaxies (SBGs) at distances~$1~\Mpc \le D < 130~\Mpc$, taken from Ref.~\cite{Lunardini:2019zcf} except that we removed the SMC and LMC (these are dwarf irregular galaxies, not SGBs, as inferred from the fact that their infrared-to-radio flux ratio is much lower than all other objects of the list), and we added the Circinus galaxy with data from the Parkes telescope ($\alpha=213.29\degree$, $\delta=-65.34\degree$, $D=4.21$, $S_{1.4~\mathrm{GHz}}=1.50~\mathrm{Jy}$); these galaxies were assigned weights proportional to their radio flux at~$1.4~\mathrm{GHz}$, see also Ref.~\cite{PierreAuger:2021rfz}.

We define the test statistic
\begin{align*}
    \TS(\psi,f,E_{\min}) &= 2 \ln \frac {L(\psi,f,E_{\min})}{L(\psi,0,E_{\min})} \,,
    & L(\psi,f,E_{\min}) &= \prod_{E_i \ge E_{\min}} \frac{\Phi(\n_i;\psi,f)\omega(\n_i)}{\int_{4\pi}\dO \Phi(\n;\psi,f)\omega(\n)} \,,
\end{align*}
where $\omega(\n)$~is the combined directional exposure of the two data sets, and the flux model is
\begin{align*}
    \Phi(\n; \psi, f) = f \Phi_\text{signal}(\n; \psi) + (1-f) \Phi_\text{background} \,,
\end{align*}
where the contribution of each source is a von Mises--Fisher distribution and the background is isotropic (normalized to~1 over the whole sphere):
\begin{align*}
    \Phi_\text{signal}(\n; \psi) &= 
    \frac{1}{\sum_j w_s}
    \sum_j w_s \frac{\psi^{-2}}{4\pi\sinh\psi^{-2}} \exp\left(\psi^{-2}\n_s \cdot \n\right) \,; & \Phi_\text{background} &= \frac{1}{4\pi} \,.
\end{align*}
Here
$E_i$~and $\n_i$~are the energy and arrival direction of the $i$-th event; $w_s$~and $\n_s$~are the weight and position of the $s$-th source candidate, and $\psi$~is the root-mean-square deflection per transverse dimension (i.e.\ the total r.m.s.\ deflection is~$\sqrt{2}\times\psi$). 
The analysis is repeated using energy thresholds of~$32~\EeV, 33~\EeV, \ldots, 80~\EeV$ on the Auger scale, corresponding to~$40.5~\EeV, 41.9~\EeV, \ldots, 106.8~\EeV$ on the TA scale.

In this work, we neglect the energy losses undergone by cosmic rays, because, given the distance distributions of the objects we are considering, the effect of energy losses on the results can be presumed to be relatively small (in the case of all galaxies) or negligible (in the case of SBGs). This enables us to be agnostic about the UHECR injection properties, thus reducing the number of model parameters for this search.

\begin{figure}[h]
\centering
\includegraphics[width=0.9\textwidth]{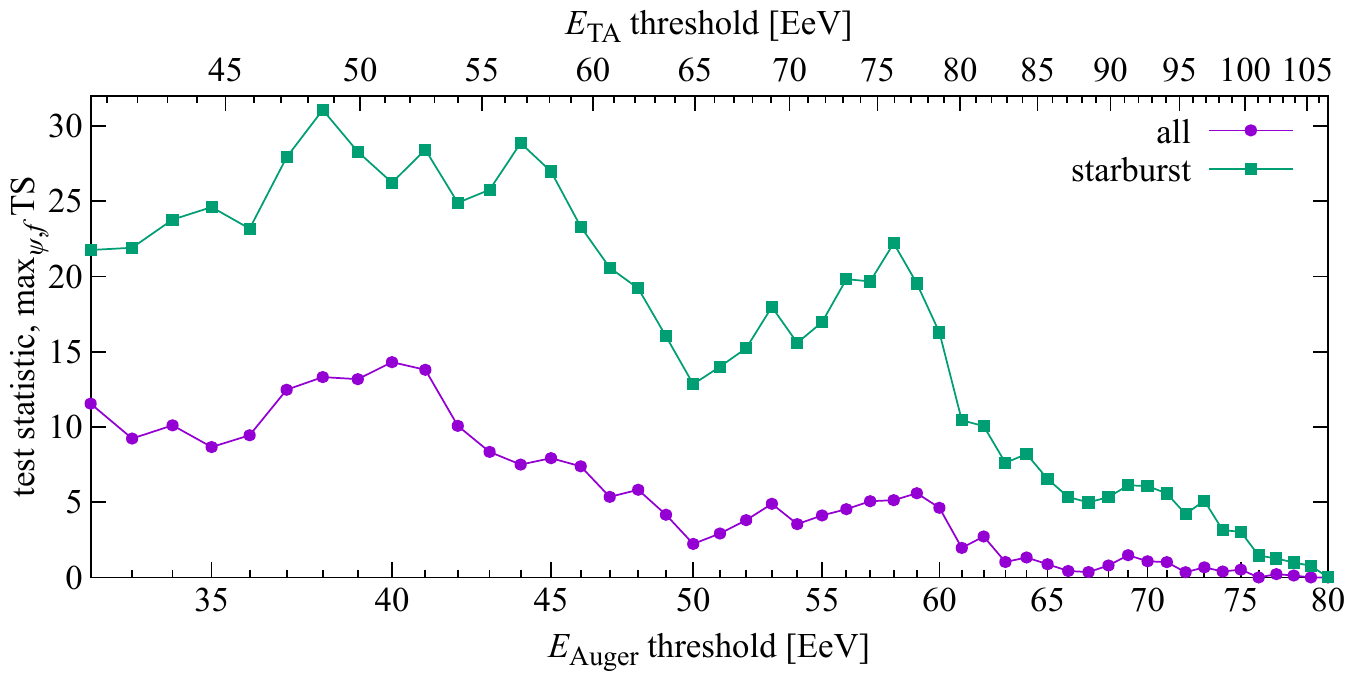}%
\caption{The TS as a function of the energy scale for the SBG catalogue (green) and the all-galaxies catalogue (violet).}
\label{sgb:fig2}
\end{figure}

\begin{figure}[h]
\centering
\includegraphics[width=0.48\textwidth]{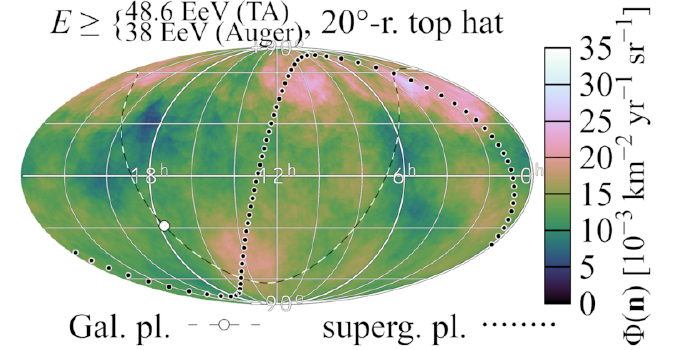}%
\hspace{0.02\textwidth}%
\includegraphics[width=0.48\textwidth]{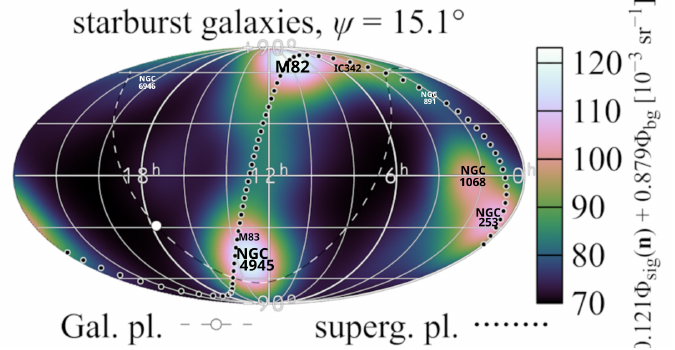}%
\caption{The observed UHECR flux above 38\,EeV on the Auger scale (49\,EeV on the TA scale) (left), compared to the starburst galaxy model predictions (right).}
\label{sgb:fig1}
\end{figure}

We present the result for the $\TS$ in figure~\ref{sgb:fig2}. The most significant correlation we found gives~$\TS = 31.1$ with the SGB catalogue, using $E_{\min} = 38\,\EeV$ on the Auger scale ($49\,\EeV$ on the TA scale),  $\psi = {15.1\degree}_{-3.0\degree}^{+4.6\degree}$, and~$f = 12.1\%_{-3.1\%}^{+4.5\%}$; the observed flux and the model prediction are compared in figure~\ref{sgb:fig1}. The post-trial significance of this is~$4.7\sigma$.\footnote{The post-trial significance is found accounting for the scans over all three parameters: $\psi$, $f$ and~$E_{\min}$.} This is a significant increase compared to the previous analysis~\cite{TelescopeArray:2021gxg}, wherein the $\TS=27.2$ corresponded to a post-trial significance of~$4.2\sigma$; the Auger-only analysis presented in~\cite{PierreAuger:2022axr} obtained $TS=25.0$ and a post-trial significance of~$4.0\sigma$. The correlation with the catalogue of all galaxies is much weaker at~$\TS=14.3$ using $E_{\min} = 40\,\EeV$ on the Auger scale ($51\,\EeV$ on the TA scale), $\psi = {28\degree}_{-12\degree}^{+11\degree}$, and~$f = 41\%_{-18\%}^{+29\%}$, with a post-trial global significance of~$2.7\sigma$---this has decreased compared to our previous study in which the $\TS=16.2$ returned a global significance of~$2.9\sigma$.

\section{Summary and Outlook}

In this work we have updated the full-sky searches for large-scale and medium-scale anisotropies in the distribution of UHECR arrival directions, making use of the latest data sets available, namely 17~years of data taken with the Pierre Auger Observatory (the same used in our previous analyses) and 14~years of data taken with the Telescope Array (three years more compared to our previous study). Using the same methods outlined in~\cite{TelescopeArray:2021gxg,TelescopeArray:2021ygq} the most important findings are: (1) a mild increase in the significance of the $d_y$ dipole component which grew from $d_y = 4.8\%\pm1.1\%\pm0.0\%$ to $d_y = 5.0\%\pm1.1\%\pm0.0\%$; (2) a marked increase in the post-trial statistical significance of the correlation with SBGs, which in this study is found to peak at~$4.7\sigma$ (cf.~$4.2\sigma$ in the previous analysis) using $E_{\min} = 38\,\EeV$ on the Auger scale ($49\,\EeV$ on the TA scale), with r.m.s.\ deflection parameter $\psi = {15.1\degree}_{-3.0\degree}^{+4.6\degree}$, and correlated flux fraction~$f = 12.1\%_{-3.1\%}^{+4.5\%}$.

Besides the obvious increase in statistics, our results complement single-hemisphere analyses in two qualitative ways. In the case of large-scale anisotropies, having full-sky coverage means that we do not need to assume anything about the UHECR flux, and the first moments of the harmonic flux decomposition can be recovered without bias. In the case of medium-scale anisotropies, we are able to observe all comparable sources; for example in the case of SBGs, the two most UHECR-bright sources, NGC4945 and M82, are in opposite hemispheres. 

The astrophysical interpretation of the UHECR--SBG association is complicated by our incomplete knowledge of intergalactic and Galactic magnetic fields as well as the UHECR mass composition. Indeed, in our analysis, in order to reduce statistical penalties, the~$\TS$ was based on a simple model that does not take into account the energy losses of UHECRs (which depend on their mass composition), coherent magnetic deflections, and the possibility of several anisotropic classes of sources at once. In order to estimate their effects, and therefore to better interpret our results, we are generating simulated sets of data based on a variety of scenarios, which we will subject to the same analyses as the observational data to test which simulations return similar values for~$\psi, f, \TS$ as the data.

TA is undergoing an upgrade (TA$\times$4) which will increase its area by a factor of~4, and which will rapidly reduce statistical uncertainties in the northern hemisphere---currently a bottleneck for our combined analyses. New scintillation and radio detectors are being added to the existing water-Cherenkov and fluorescence detectors (the AugerPrime upgrade), which will reduce statistical and systematic uncertainties on UHECR masses. Moreover, improved mass estimation from new analysis techniques (e.g., machine learning) are being developed~\cite{TelescopeArray:2018bep,PierreAuger:2021fkf}. The combination of these will allow us to study mass-dependent anisotropies, potentially allowing us to disentangle the effects of magnetic deflections from the distribution of UHECR sources, for instance by selecting high-rigidity samples with smaller magnetic deflections. To conclude, given the statistical significance of the results presented here, in particular the~$4.7\sigma$ correlation of UHECRs with the SBG catalogue, it is imperative that the two observatories continue running through this decade in order to decidedly confirm this correlation.

\newpage
\bibliography{biblio}

\begin{thebibliography}{21}

\bibitem{AlvesBatista:2021gzc}
R.~Alves~Batista et~al. (2021), \texttt{2110.10074}

\bibitem{Anchordoqui:2018qom}
L.A. Anchordoqui, Phys. Rept. \textbf{801}, 1 (2019), \texttt{1807.09645}

\bibitem{Abbasi:2016kgr}
R.U. Abbasi et~al., Astropart. Phys. \textbf{86}, 21 (2017),
  \texttt{1608.06306}

\bibitem{PierreAuger:2017pzq}
A.~Aab et~al. (Pierre Auger), Science \textbf{357}, 1266 (2017),
  \texttt{1709.07321}

\bibitem{PierreAuger:2021dqp}
R.~de~Almeida et~al. (Pierre Auger), PoS \textbf{ICRC2021}, 335 (2021)

\bibitem{TelescopeArray:2021gxg}
A.~di~Matteo et~al. (Telescope Array, Pierre Auger), PoS \textbf{ICRC2021}, 308
  (2021)

\bibitem{PierreAuger:2018qvk}
A.~Aab et~al. (Pierre Auger), Astrophys. J. Lett. \textbf{853}, L29 (2018),
  \texttt{1801.06160}

\bibitem{PierreAuger:2021rfz}
P.~Abreu et~al. (Pierre Auger), PoS \textbf{ICRC2021}, 307 (2021)

\bibitem{PierreAuger:2015eyc}
A.~Aab et~al. (Pierre Auger), Nucl. Instrum. Meth. A \textbf{798}, 172 (2015),
  \texttt{1502.01323}

\bibitem{PierreAuger:2022axr}
P.~Abreu et~al. (Pierre Auger), Astrophys. J. \textbf{935}, 170 (2022),
  \texttt{2206.13492}

\bibitem{TelescopeArray:2012uws}
T.~Abu-Zayyad et~al. (Telescope Array), Nucl. Instrum. Meth. A \textbf{689}, 87
  (2013), \texttt{1201.4964}

\bibitem{TelescopeArray:2021ygq}
P.~Tinyakov et~al. (Telescope Array, Pierre Auger), PoS \textbf{ICRC2021}, 375
  (2021)

\bibitem{TelescopeArray:2014ahm}
A.~Aab et~al. (Telescope Array, Pierre Auger), Astrophys. J. \textbf{794}, 172
  (2014), \texttt{1409.3128}

\bibitem{diMatteo:2018vmr}
A.~di~Matteo, O.~Deligny, K.~Kawata, R.M. de~Almeida, M.~Mostaf\'a,
  E.~Moura~Santos, H.~Sagawa, P.~Tinyakov, I.~Tkachev, N.~Toshiyuki, JPS Conf.
  Proc. \textbf{19}, 011020 (2018)

\bibitem{PierreAuger:2019oxg}
J.~Biteau et~al. (Pierre Auger, Telescope Array), EPJ Web Conf. \textbf{210},
  01005 (2019), \texttt{1905.04188}

\bibitem{Eichmann:2020adn}
B.~Eichmann, T.~Winchen, JCAP \textbf{04}, 047 (2020), \texttt{2001.01530}

\bibitem{diMatteo:2017dtg}
A.~di~Matteo, P.~Tinyakov, Mon. Not. Roy. Astron. Soc. \textbf{476}, 715
  (2018), \texttt{1706.02534}

\bibitem{Ding:2021emg}
C.~Ding, N.~Globus, G.R. Farrar, Astrophys. J. Lett. \textbf{913}, L13 (2021),
  \texttt{2101.04564}

\bibitem{Lunardini:2019zcf}
C.~Lunardini, G.S. Vance, K.L. Emig, R.A. Windhorst, JCAP \textbf{10}, 073
  (2019), \texttt{1902.09663}

\bibitem{TelescopeArray:2018bep}
R.U. Abbasi et~al. (Telescope Array), Phys. Rev. D \textbf{99}, 022002 (2019),
  \texttt{1808.03680}

\bibitem{PierreAuger:2021fkf}
A.~Aab et~al. (Pierre Auger), JINST \textbf{16}, P07019 (2021),
  \texttt{2101.02946}

\end{thebibliography}

\end{document}